\documentclass[aps,prb,reprint,groupedaddress,showpacs]{revtex4-1}
\usepackage{graphicx}
\usepackage{mathrsfs} 
\usepackage{amsmath,amssymb}
\usepackage{bm}
\newcommand{\1}{\mbox{1}\hspace{-0.25em}\mbox{l}} %

\begin{document}


\title{Determination of Boundary Scattering, Intermagnon Scattering, and \\ 
the Haldane Gap in Heisenberg Chains}


\author{Hiroshi \textsc{Ueda}}
\email[]{ueda@aquarius.mp.es.osaka-u.ac.jp}
\author{Koichi \textsc{Kusakabe}}
\email[]{kabe@mp.es.osaka-u.ac.jp}
\affiliation{Graduate School of Engineering Science, Osaka University, 
Toyonaka, Osaka 560-8531, Japan}


\date{\today}

\begin{abstract}
Low-lying magnon dispersion in a $S=1$ Heisenberg antiferromagnetic (AF) chain is analyzed using the non-Abelian DMRG method. The scattering length $a_{\rm b}$ of the boundary coupling and the inter-magnon scattering length $a$ are determined. The scattering length $a_{\rm b}$ is found to exhibit a characteristic diverging behavior at the crossover point. In contrast, the Haldane gap $\Delta$, the magnon velocity $v$, and $a$ remain constant at the crossover. Our method allowed estimation of the gap of the $S=2$ AF chain to be $\Delta = 0.0891623(9)$ using a chain length longer than the correlation length $\xi$.

\end{abstract}

\pacs{75.10.Jm, 75.40.Mg}

\maketitle
\section{Introduction}
To form a better understanding of interacting many-body systems, it is very important to determine an effective field theory and to clarify the low-energy physics involved. 
In the physics of low-dimensional quantum systems, 
considerable attention has been paid to 
the one-dimensional antiferromagnetic (AF) integer-spin Heisenberg model 
following the discovery of the Haldane gap.~\cite{Haldane1983464,PhysRevLett.50.1153} 
Precise determination of the gap has been 
reported by several authors.~\cite{PhysRevB.48.3844,PhysRevLett.87.047203,JPSJ.78.014003,PhysRevB.77.134437,1103.2286} 
Its massive elementary excitation, {\it i.e.,} the magnon, has a relativistic dispersion relation, 
which is often described by a non-linear sigma model 
(NLSM).~\cite{Haldane1983464, PhysRevLett.50.1153, Nucl_Phys_b257_397, 0953-8984-1-19-001}   

In particular, the $S=1$ AF Heisenberg chain has been widely studied both theoretically and experimentally.  
When open boundary conditions (OBC) are applied to a $S=1$ AF chain,  
owing to the unique effective $S=1/2$ spins at the ends, quasi-degeneracy appears between the 
singlet ground state and a low-lying triplet state.~\cite{PhysRevB.45.304} 
Various attempts at boundary tuning, ~\cite{PhysRevB.48.3844} as exemplified by 
attachment of real $S=1/2$ spins to maintain the high accuracy of 
the density matrix renormalization group (DMRG) method,~\cite{White:PRL69, White:PRB48} 
have shown that deformation of the boundary conditions can selectively modify the magnon wavefunction while maintaining the uniformity of the ground state.~\cite{JPSJ.78.014001} 

To form a better understanding of the physics involved in a finite chain under OBC, 
we can use the NLSM to describe the low-lying energy dispersion. 
Lou {\it et al.} have proposed usage of a form, $\sqrt{\Delta^2+v^2 \sin^2 k_{\rm eff}}$, 
for low-lying magnon dispersions, where $k_{\rm eff}$ is the effective wavenumber.~\cite{PhysRevB.62.3786}  
Here, $\Delta$ denotes the Haldane gap and $v$ is the velocity of the quasi particle. 
They described the asymptotic effects of boundary scattering and inter-magnon interactions 
in terms of the scattering lengths, $a_{\rm b}$ and $a$, which appear in $k_{\rm eff}$. 
When boundary tuning is applied by introducing an antiferromagnetic 
coupling $J_{\rm end}$ between the $S=1$ spin chain and the extra real $S=1/2$ spin, 
these scattering lengths might be effected. 
This idea motivated us to study low-lying elementary excitations 
using both the DMRG and NLSM methods by describing the bulk properties and the boundary scattering effects in terms of an effective theory. 
In this work, using the DMRG method, the energy dispersion of 
various magnon modes was determined for $S=1$ Heisenberg systems with up to 2048 spins. 
Finite-size scaling analysis was performed to determine 
the boundary scattering length and the inter-magnon scattering length, 
in addition to $\Delta$ and $v$ in the thermodynamic limit. 
We used a relation of the correlation length $\xi \sim v/\Delta$, which is known to hold approximately in the integer-spin AF Heisenberg chain.~\cite{PhysRevLett.71.1633, PhysRevB.56.R14251}
We found that $a_{\rm b}$ changed sign around a critical value of $J_{\rm end}$. 
This value should be identical to that required to make local quantities such as the local bond energy of the ground state and the spin density of long-wavelength magnons  uniform.~\cite{PhysRevB.48.3844, PhysRevB.54.4038, PhysRevB.77.134437, 0812.4513} 
In addition, a divergence-like behavior of $a_{\rm b}$ was detected 
around this critical value denoted as $J_{\rm end}^{\rm c}$. However, 
the inter-magnon scattering length was found to be constant at $a=-0.383(6)\xi$ irrespective of $J_{\rm end}$. 
In this derivation, $\Delta$, $v$, and $\xi$ were confirmed to be always independent of $J_{\rm end}$ 
in the thermodynamic limit. This allows the low-lying elementary excitations to be effectively described. 
The results indicated the presence of both itinerating magnons (IMs) and boundary magnons (BMs) bound at the ends.
At $J_{\rm end}^c$, the diagonal magnetization induced by an IM shows 
a flat structure around the center of the system when $L\gg \xi$, with $L$ being the number of $S=1$ spins. 
Both the diverging behavior of $a_{\rm b}$ and the uniform distribution of the long-wavelength magnons 
confirm the realization of bulk characteristics in an elementary excitation at 
the critical point $J_{\rm end}^{\rm c}$, where the ground state also 
has a uniform nature around the center of the system. 

Furthermore, this work clearly resolves the problem pointed out by Todo and Kato~\cite{PhysRevLett.87.047203}; there is disagreement between the DMRG~\cite{PhysRevB.60.14529} and quantum Monte Carlo (QMC) simulation results~\cite{PhysRevLett.87.047203} with respect to estimation of the excitation gap in the $S=2$ AF Heisenberg model.
The reason for this disagreement might be an inappropriate scaling assumption in the DMRG study.
This work applies finite-size scaling analysis to the excitation gap in the $S=2$ AF chain, and shows for the first time  that the corrected gap is within the error bar of the QMC value.

\section{Effective Hamiltonian}
We consider a $S=1$ AF chain with boundary $S=1/2$ spins ${\bf s}_{j}$ with $j=0$ or $L+1$, 
which is described by the following Hamiltonian. 
\begin{equation}
H(J_{\rm end}) 
= \sum_{i=1}^{L-1} {\bf S}_{i} \cdot {\bf S}_{i+1}
+ J_{\rm end}({\bf s}_{0} \cdot {\bf S}_{1} + {\bf S}_{L} \cdot {\bf s}_{L+1}), 
\label{ham1}
\end{equation}
where ${\bf S}_{i}$ represents the $S=1$ operator at the $i$-th site. 

The low-energy physics of the Hamiltonian in Eq. (\ref{ham1}) can be understood 
by using an approximate mapping onto the 
NLSM.~\cite{Haldane1983464, PhysRevLett.50.1153, Nucl_Phys_b257_397, 0953-8984-1-19-001}
We let $L\rightarrow \infty$, keeping $X=Lb$ constant, with $b$ being the lattice spacing. 
Taking into account the effective $S=1/2$ boundary modes ${\bf s}^{\rm eff}_j$, 
we obtain the following expression, 
\begin{eqnarray}
H_{\rm eff} &=& H_{\rm NLSM} 
+ \lambda_s \left[ \bm{\phi}(0) \cdot {\bf s}^{\rm eff}_{1} 
+(-1)^L \bm{\phi}(X) \cdot {\bf s}^{\rm eff}_{L} \right] \nonumber \\
&&
+ \lambda_u \left[ {\bf l}(0) \cdot {\bf s}^{\rm eff}_{1} 
+{\bf l}(X) \cdot {\bf s}^{\rm eff}_{L} \right] \nonumber \\
&&
+ \lambda_s' \left[ \bm{\phi}(0) \cdot {\bf s}_{0} 
+(-1)^L \bm{\phi}(X) \cdot {\bf s}_{L+1} \right] \nonumber \\
&&
+ \lambda_u' \left[ {\bf l}(0) \cdot {\bf s}_{0} 
+{\bf l}(X) \cdot {\bf s}_{L+1} \right] \nonumber \\
&&
+ J^{\rm eff}_{\rm end} \left[ {\bf s}^{\rm eff}_{1} \cdot {\bf s}_{0} 
+ {\bf s}^{\rm eff}_{L}\cdot {\bf s}_{L+1}\right],
\end{eqnarray}
with the bulk part of the NLSM expressed as \begin{equation}
H_{{\rm NLSM}} = \frac{v}{2} \int_{0}^{X} 
dx \left[ g {\bf l}^2+ \frac{1}{g} \left( \frac{\partial \bm{\phi}}{\partial x} \right)^2 \right],
\end{equation}
where $\bm{\phi}$ and ${\bf l} \equiv (1/vg)\bm{\phi} \times \partial_t\bm{\phi}$ 
are low-energy Fourier modes of the spin operators with wave vectors near $\pi$ and 0. 
The coupling parameter and the velocity are given as $g=\frac{2}{S}$, $v = 2S$. 
Since all the bare couplings are antiferromagnetic, 
solutions for the bulk fields follow the 
Neumann boundary conditions (NBC) : $d\bm{\phi}/dx|_{x=0,X}=0$.~\cite{PhysRevB.62.3786} 
The $\lambda_u$ and $\lambda_u'$ terms 
produce an effective boundary repulsive potential on an IM, 
and $J_{\rm end}^{\rm eff}$ is a renormalized coupling constant. 

\begin{figure}
\begin{center}
\includegraphics[width=80mm]{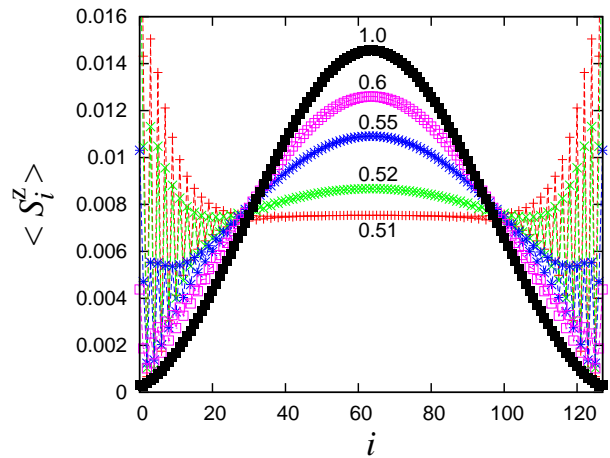}
\end{center}
\caption{(Color online) Distribution of local magnetization $\langle S_j^z \rangle$ 
for a single magnon state with $S_{\rm tot}=1$ for various values of $J_{\rm end}$.
}
\label{sz_dis}
\end{figure}

The validity of this description is also confirmed by examining the spin density of an IM shown in Fig. \ref{sz_dis}. 
When $J_{\rm end}$ is larger than $J_{\rm end}^{\rm c}$, the lowest triplet mode has 
itinerating behavior. Indeed, we see that $\langle S_{i}^z \rangle$ exhibits a cosine-like behavior 
for $J_{\rm end}=1.0$ owing to both strong repulsive coupling 
via $\lambda_u$ and $\lambda_u'$, and the NBC on $\bm{\phi}(x)$.  
When $J_{\rm end}$ approaches $J_{\rm end}^c \sim 0.51$, 
the IM mode becomes uniform around the center of the chain but 
$\langle S_{i}^z \rangle$ exhibits damped oscillations near the two ends.  
This known solution suggests that the mode should continuously change into 
an end mode ${\bf s}^{\rm eff}_j$ 
in the low-energy eigenstate when $J_{\rm eff}<J_{\rm eff}^c$. 

Thus, the dispersion relation for $N$ itinerating magnon modes 
at low energies in the dilute limit 
may be simply reproduced by a nonrelativistic effective Hamiltonian for $N$ virtual particles, 
\begin{eqnarray}
H_{\rm eff}(J_{\rm end}) & = &
\frac{1}{2m}\sum^{N}_{i=1}\frac{d^2}{dx^{2}_{i}} + \sum_{\langle i,j \rangle} V(x_i-x_j) 
\nonumber \\
&+&
\sum^{N}_{i=1}[V_{\rm b}(J_{\rm end}, x_i) + V_{b}(J_{\rm end}, X-x_i)], 
\label{ham_eq}
\end{eqnarray}
where $0 \leq x_j \leq X$, with a wavefunction obeying the Neumann boundary condition: 
$\partial_{j} \psi(x_1,\cdots, x_N)|_{x_j=0,X} = 0$. 
Here we use the Einstein relation $m \equiv \Delta/v^2$. 
Effective short-range interactions between IMs 
and between an IM and a BM  are represented by $V$ and $V_{\rm b}$, respectively. 
We expect they are short-range function with range of the order of the correlation length $\xi$. 
All of the effects of $J_{\rm end}$ are produced by the boundary potential $V_{\rm b}(J_{\rm end},x)$
In the asymptotic region, the effects of $V$ and $V_{\rm b}$ 
appear as scattering phase shifts, which are represented by $a$ and $a_{\rm b}$. 

We now identify low-lying magnon modes. Each mode is specified by a total spin of $S_{\rm tot}$. 
When $J_{\rm end}$ is small and positive, since we have 
two effective $S=1/2$ spins creating the bulk low-lying triplet 
and two real $S=1/2$ spins, we need to polarize these four spins before we can create 
one IM. 
In this case, the effective chain length for the IM becomes $L-2a_{\rm b}$ 
and $k_{\rm eff}=\pi/(L-2a_{\rm b})$, 
when the system is about two times longer than the correlation length $\xi$. 
Therefore, we have the relation: 
\begin{equation}
E_{32} = \sqrt{\Delta^2 + v^2\sin^2\frac{\pi}{L-2a_{\rm b}}},
\label{E32}
\end{equation}
where $E_{ji}=E_{j}-E_{i}$ and $E_{j}$ and $E_{i}$ are the lowest energy of the $S_{\rm tot}=j$ and $S_{\rm tot}=i$ states. 
The energy spectrum $E_{42}$ for two IMs is given by 
\begin{equation}
E_{42} 
= \sum_{j=1}^2 \sqrt{\Delta^2 + v^2\sin^2\frac{j\pi}{L-2a_{\rm b}-a}},
\label{E42}
\end{equation}
where we use the small-$k$ approximation for the magnon-magnon phase shift.
When $J_{\rm end}$ becomes large enough, the effective boundary $S=1/2$ modes couple 
strongly with the real $S=1/2$ spins. 
In this condition, 
the low-lying magnon states are IMs, and the formulas for $E_{10}$ and $E_{20}$ are, respectively, similar to Eq. (\ref{E32}) and Eq. (\ref{E42}). 
Thus, we can conclude that a crossover value of $J_{\rm end}^{\rm c}$ exists, where the low energy spectrum changes qualitatively. 

\section{Numerical results}
We used the non-Abelian DMRG method (NA-DMRG)~\cite{Mc_non-abe} to estimate
the energy spectrum of the lowest $S_{\rm tot}=0, 1, 2, 3$, and 4 states for finite systems. 
Numerical convergence during finite system sweeping was accelerated by the use of a wave function prediction method.~\cite{PhysRevLett.77.3633, JPSJ.64.4084, JPSJ.75.014003, JPSJ.77.114002, 0804.2509, JPSJ.79.044001}
Since the number of kept states for the block spin is up to $m_s = 512$, the truncation error is 
smaller than $1.0 \times 10^{12}$ in the lowest $S_{\rm tot} = 4$ state. 
This corresponds to a number of kept states of $m_{s^{\rm z}} \sim 2500 - 2700$ in the standard DMRG. 
In this case, the numerical cost of the standard DMRG is about 110 - 140 times higher than that of 
NA-DMRG, because in the DMRG it varies as the cube of the number of kept states. 
The system size $L+2$ is up to 2048, where the two extra spins indicate the boundary $S=1/2$ spins. 

\begin{figure}
\begin{center}
\includegraphics[width=80mm]{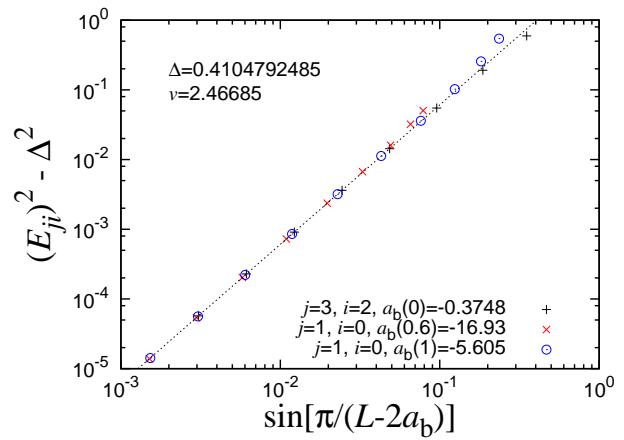}
\end{center}
\caption{(Color online) Single magnon energy with $J_{\rm end}$ under the condition $m_s=512~(m_{s^{\rm z}} \sim 2500)$. The dotted line represents $v^2 \sin^2(\pi/(L-2a_{\rm b}))$. 
}
\label{sm}
\end{figure}

The energy of a single IM as a function of the system size is shown in Fig. \ref{sm}. 
The target energy spectrum is $E_{32}$ when $J_{\rm end} = 0$, and $E_{10}$ when $J_{\rm end} = 0.6$ or 1. 
To estimate $\Delta$, $v$ and $a_{\rm b}$, we generated sequences $A^{*}(L_{0}+2)$ for different values of $L=L_{0}$, where $A^{*}(L)$ denotes finite values 
of $A = \Delta, v$ and $a_{\rm b}$ in the thermodynamic limit. 
The sequences were determined  by least square fitting with the function $\sqrt{\Delta^2 + v^2\sin^2\frac{\pi}{L-2a_{\rm b}}}$ 
for IM energies of $L+2=2^\ell( L_{0}+2 )$, where $\ell=0,\pm1$. 
The value of $A$ was estimated by power-law extrapolation with elements of $A^{*}(512)$ and $A^{*}(1024)$. 
The estimation error was taken to be $|A - A^{*}(1024)|$. 
Based on the optimum boundary scattering length $a_{\rm b}(J_{\rm end})$ for each $J_{\rm end}$, 
we found a universal finite size dependence for a fixed energy gap $\Delta$ and spin velocity $v$. 
As a result, we showed that 
only the boundary scattering length $a_{\rm b}$ was affected by changing $J_{\rm end}$, whereas $\Delta$ and $v$ 
were independent of $J_{\rm end}$ (See Table \ref{tab1}). 
This result is consistent with the effective model in Eq. (\ref{ham_eq}). 

\begin{figure}
\begin{center}
\includegraphics[width=80mm]{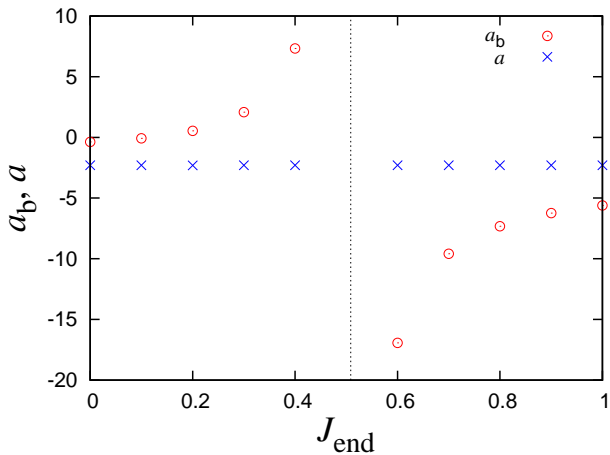}
\end{center}
\caption{(Color online) Boundary scattering length and intermagnon scattering length as a function of $J_{\rm end}$. 
The dotted line represents $J_{\rm end} = 0.50865$. 
}
\label{stm}
\end{figure}

The estimated values of $\Delta$, $v$, and $\xi=v/\Delta$ are consistent to within $\Delta=0.4104792485(4)$, 
$v=2.46685(2)$ and $\xi = 6.00967(5)$, respectively, except for a somewhat larger error at 
$J_{\rm end}=0.4$ and $0.6$, which are closest to $J_{\rm end}^{\rm c}$.  
Since our data is obtained by extrapolation using system sizes larger than those treated in former 
studies,~\cite{PhysRevB.48.3844,PhysRevLett.87.047203}  
our results show meaningful differences. 
The reported value of $a_{\rm b}(J_{\rm end}=0) = -1$ in Ref.~[\cite{PhysRevB.62.3786}] is 
about three times larger than our result of $a_{\rm b}(J_{\rm end}=0) = -0.3748(1)$. 
The value of $a_{\rm b}(J_{\rm end})$ changes rather dramatically with $J_{\rm end}$, (Fig.~\ref{stm}) with a change in sign even occurring around $J_{\rm end} \sim 0.5$. 
The values seem to diverge around $J_{\rm end}^{\rm c}$. 
When the boundary scattering length becomes $a_{\rm b} \rightarrow -\infty$, 
$k_{\rm eff}=\pi/(L-2a_{\rm b})$ approaches zero, 
and the energy of the lowest IM is almost at its minimum value, $\Delta$, and is independent of 
$J_{\rm end}$. 
This is consistent with the report in Ref.~[\cite{PhysRevB.48.3844, PhysRevB.54.4038}].
However, we should note that the above picture holds only when $L\gg a_{\rm b}$, 
requiring a high performance simulation tool such as NA-DMRG. 

In the same manner, using the estimated $\Delta$, $v$, and $a_{\rm b}$, 
we determined the inter-magnon scattering length $a$. 
The target energy spectrum is $E_{42}$ when $J_{\rm end} < J_{\rm end}^{\rm c}$, 
and is $E_{20}$ when $J_{\rm end} > J_{\rm end}^{\rm c}$. 
With a common $a$, universal behavior is observed in the large $L$ region. 
The estimated values of $a$ are consistent to within  $a=-2.30(4)=-0.383(6)\xi$ except for 
a somewhat larger error at $J_{\rm end}=0.4$ and 0.6. (See Table \ref{tab1} and Fig. \ref{stm}.) 
Thus, we conclude that the value of $a$ is independent of $J_{\rm end}$. 
The estimated value of $a$ is comparable to $a=-0.32\xi$ in Ref.~[\cite{PhysRevB.62.3786}]. 
In Fig. \ref{stm}, the dotted line represents $J_{\rm end} = 0.50865$ 
determined in Ref.~[\cite{PhysRevB.77.134437}]. 

\begin{table*}
\begin{center}
\caption{Results of numerical simulations for a single itinerating magnon, showing magnon energy $\Delta$, magnon velocity $v$, boundary scattering length $a_{\rm b}$, intermagnon scattering length $a$, and correlation length $\xi=v/\Delta$, $a_{\rm b}/\xi$ and $a/\xi$.}
\label{tab1}
\begin{ruledtabular}
\begin{tabular}{ c l l l l l l l }												
$J_{\rm end}$	&	$\Delta$	&	$v$	&	$a^{}_{\rm b}$	&	$a$	&	$\xi^{}$	&	$a^{}_{\rm b}$/$\xi^{}$	& $a^{}_{}$/$\xi^{}$ \\
\hline													
0	&	0.4104792487(1)	&	2.466838(1)	&	~-0.3748(1)	& ~-2.30(2)	& 6.009654(1)	&	~-0.06237(2)	&~-0.383(4)	\\
0.1	&	0.4104792486(1)&	2.466844(2)	&	~-0.0836(3)	& ~-2.301(2) &	6.009669(4)	&	~-0.01391(5) &	~-0.3830(4) \\
0.2	&	0.4104792487(1)	&	2.46684(1)	&	~~0.540(2)	& ~-2.303(5)	& 6.00966(3)	&	~~0.0898(3)	& ~-0.3833(9) \\
0.3	&	0.4104792486(4)	&	2.46684(4)	&	~~2.081(8)	& ~-2.30(4)	& 6.0096(1)	&	~~0.346(1)	& ~-0.384(6) \\
0.4	&   0.410479248(2)	&	2.4668(3)	&	~~7.33(5)	& ~-2.3(2)	& 6.0098(7)	&	~~1.220(9) & ~-0.38(5) \\ 
\hline
0.6	&   0.410479248(2)	&	2.4668(2)	&	~-16.93(2)	&	~-2.3(2)	&	6.0096(5)	&	~-2.821(6)	& ~-0.38(3) \\
0.7 &   0.4104792483(2)	&	2.46685(3)	&	~-9.586(5)	&	~-2.30(3)	&	6.00968(7)	&	~-1.5951(9)	&	~-0.382(4)	\\
0.8	&	0.4104792483(2)	&	2.46685(2)	&	~-7.317(3)	&	~-2.30(2)	&	6.00968(4)	&	~-1.2176(5)	&	~-0.383(3)	\\
0.9	&	0.4104792483(2)	&	2.46685(2)	&	~-6.233(3)	&	~-2.30(2)	&	6.00968(4)	&	~-1.0372(5)	&	~-0.383(3)	\\
1.0	&	0.4104792485(1)	&	2.46684(1)	&	~-5.605(3)	&	~-2.30(2)	&	6.00967(4)	&	~-0.9328(5)	&	~-0.383(3)	\\
\end{tabular}
\end{ruledtabular}
\end{center}
\end{table*}

\section{Application to $S=2$ Heisenberg Chain and Conclusions} 
We have shown that 
the energy spectrum modified by the tuning parameter $J_{\rm end}$ can be fitted 
using an effective massive relativistic dispersion 
with a boundary scattering length $a_{\rm b}(J_{\rm end})$ modified for lattice models. 
The intermagnon scattering length $a$ is constant irrespective of $J_{\rm end}$, as well as
other bulk quantities including the Haldane gap, the magnon velocity, and the correlation length. 
In contrast, $a_{\rm b}(J_{\rm end})$ drastically changes around $J_{\rm end} \sim 0.5$, 
representing a crossover point for the physics at the boundary. 

Analysis of the boundary scattering length and intermagnon scattering length was also carried out for a
$S=2$ AF Heisenberg chain, 
where ${\bf S}_i$ and ${\bf s}_i$ in the Hamiltonian in eq. (\ref{ham1}) represent the $S = 2$ and $S=1$ operators, respectively.
In addition, we choose $J_{\rm end}=1$, so that the low-lying magnon states are IMs, and a similar formula for $E_{10}$ is obtained to that shown in Eq. (\ref{E32}).
Our data was taken using $m_{s}=1024$, which corresponds to $m_{s^{\rm z}} \sim 6000$, 
and large systems up to $L+2=2048$, 
The truncation error is smaller than $1 \times 10^{-11}$. 
Note that the numerical cost using NA-DMRG is about 200 times less than that for the standard DMRG in this case. 
In contrast to a former report,~\cite{PhysRevB.62.13832} 
our results suggests a large value of $a_{\rm b}(J_{\rm end}=1)=-33(1)=-0.67(2)\xi$. 
Our calculations give the excitation gap $\Delta = 0.0891623(9)$, the spin velocity $v=4.42(1)$, 
and the correlation length $\xi=49.6(1)$. In particular, the value of $\Delta$ obtained in the present study is consistent with the value of $0.08917(4)$ determined by quantum Monte Carlo simulations.~\cite{PhysRevLett.87.047203}
The estimate of the Haldane gap has thus been improved by two more significant digits. 
This indicates the ability of the effective theory to correctly describe the low-energy physics, and the usefulness of the proposed numerical approach is studying such problems

It would be of interest to apply the approach used in this work to finite size scaling with different boundary tuning methods such as hyperbolic deformation.~\cite{JPSJ.78.014001,PTP.124.389,1102.0845}. 
In such a situation, the excited quasi particle is weakly confined near the center of the system under the deformation. 
In Ref.~[\cite{PTP.124.389}], we showed that it is necessary to introduce an additional parameter $d$ and replace $L + 1$ by $L + d$ in order to reduce higher-order corrections. 
This replacement is introduced in the effective model shown in Eq. (\ref{ham_eq})
by considering the effective boundary scattering. 
The boundary scattering length has an important and universal influence on excitation energy scaling as long as there are chain ends. 

In this work, a relation $\xi = v/\Delta$ is used to estimate the correlation length in each spin-$S$ chain. 
If we use an assumption for a relation between the low-energy dispersion curve and the ground state correlation length, namely $\sinh \xi^{-1} = \Delta/v$ in this case,~\cite{PhysRevB.64.104432} the correlation lengths are evaluated as $6.03720(9)$ in $S=1$ and $49.6(1)$ in $S=2$. 
In the case of $S=1$, we have a meaningful different value from the former estimation.
On the contrary, the difference is not confirmed in the case of $S=2$. 
To find correct relation between the low-energy dispersion and the correlation length in each spin-$S$ AF Heisenberg chain is a future issue. 

For a final development of the low-lying effective field theory to describe the low-lying magnon dispersions, 
discussions for rigorous results of wave functions and energy dispersions for low-lying states are important. 
The effective dispersion relation of $\sqrt{\Delta^2 + v^2 \sin^2 k_{\rm eff}}$ is known to 
appear in the Haldane phase~\cite{PRL.69.3571} and also 
in the massive phase of the $S=1/2$ XXZ model.~\cite{Takahashi} 
The Bethe-ansatz solutions for OBC 
suggest that an analogous crossover from an IM with real $k_{\rm eff}$ to 
a BM with a damping nature can be found as a continuous change from a real to an imaginary rapidity.~\cite{Deguchi_Yue_Kusakabe} 
\begin{acknowledgments}
This work was supported in part by a Grant-in-Aid for JSPS Fellows, Grant-in-Aids (No. 19051016), Global COE Program (Core Research and Engineering of Advanced Materials - Interdisciplinary Education Center for Materials Science) from the Ministry of Education, Culture, Sports, Science and Technology of Japan.

\end{acknowledgments}

%

\end{document}